\documentstyle[preprint,eqsecnum,aps,epsf]{revtex}

\tightenlines
\begin{document}
\preprint{\vbox{
\hbox{INPP-UVA-97-08} 
\hbox{December, 1997} 
\hbox{hep-ph/9802206}
}}
\draft
\def\be{\begin{eqnarray}}
\def\en{\end{eqnarray}}
\def\non{\nonumber}
\def\la{\langle}
\def\ra{\rangle}
\def\up{\uparrow}
\def\dw{\downarrow}
\def\ep{\varepsilon}
\def\ms{\overline{\rm MS}}
\def\ums{{\mu}_{_{\overline{\rm MS}}}}
\def\u{\mu_{\rm fact}}
\def\pr{{\sl Phys. Rev.}~}
\def\ijmp{{\sl Int. J. Mod. Phys.}~}
\def\jp{{\sl J. Phys.}~}
\def\mpl{{\sl Mod. Phys. Lett.}~}
\def\prp{{\sl Phys. Rep.}~}
\def\prl{{\sl Phys. Rev. Lett.}~}
\def\pl{{\sl Phys. Lett.}~}
\def\np{{\sl Nucl. Phys.}~}
\def\ppnp{{\sl Prog. Part. Nucl. Phys.}~}
\def\zp{{\sl Z. Phys.}~}

\title{\bf Orbital Angular Momentum in Chiral Quark Model\\ }

\author{Xiaotong Song\\}

\address{Institute of Nuclear and Particle Physics\\
Jesse W. Beams Laboratory of Physics\\
Department of Physics, University of Virginia\\
Charlottesville, VA 22901, USA\\}
\date{December 23, 1997, Revised Jan., March, and April 1998}
\maketitle
\vskip 12pt
\begin{abstract}
We developed a new and unified scheme for describing both quark 
spin and orbital angular momenta in symmetry-breaking chiral 
quark model. The loss of quark spin in the chiral splitting 
processes is compensated by the gain of the orbital angular 
momentum carried by quarks and antiquarks. The sum of both 
spin and orbital angular momenta carried by quarks and antiquarks 
is 1/2. The analytic and numerical results for the spin and orbital 
angular momenta carried by quarks and antiquarks in the nucleon are 
given. Extension to other octet and decuplet baryons is also 
presented. Possible modification and application are discussed.
\end{abstract}

\bigskip
\bigskip
\bigskip

\pacs{13.88.+e,~11.30.Hv,~12.39.Fe,~14.20.Dh\\}

\newpage

\leftline{\bf I. Introduction}

One of important tasks in hadron physics is to reveal the
internal structure of the nucleon. This includes the study
of flavor, spin and orbital components shared by the quarks
and gluons in the nucleon. These structures are intimately 
related to the nucleon properties : spin, magnetic moments, 
axial coupling constant, elastic form factors, and the deep 
inelastic structure functions. The polarized deep-inelastic 
scattering (DIS) data \cite{emc,smc-slac-hermes,smc97} indicate 
that the quark spin only contributes about one third of the 
nucleon spin or even less. A natural and interesting question 
is where is the {\it missing} spin ? Intuitively and also from 
the quantum chromodynamics (QCD) \cite{ji-jm}, the nucleon 
spin can be decomposed into the quark and gluon pieces
$$
{1\over 2}=<J_z>_{q+\bar q}+<J_z>_{G}={1\over 2}\Delta\Sigma+
<L_z>_{q+\bar q}+<J_z>_{G}
\eqno (1)
$$
Without loss of generality, in (1) the proton has been chosen to 
be {\it longitudinal polarized} in the $z$ direction and has helicity 
of $+{1\over 2}$. The angular momentum $<J_z>_{q+\bar q}$ has been
decomposed into the spin and orbital pieces in (1). The total spin from 
quarks and antiquarks is ${1\over 2}\Delta\Sigma={1\over 2}\sum\limits
[\Delta q+\Delta\bar q]=<s_z>_{q+\bar q}$, where 
$\Delta q\equiv q_{\up}-q_{\dw}$, and 
$\Delta{\bar q}\equiv{\bar q}_{\up}-{\bar q}_{\dw}$, and $q_{\up,\dw}$ 
(${\bar q}_{\up,\dw}$) are quark (antiquark) {\it numbers} of spin 
parallel and antiparallel to the nucleon spin, or more precisely, quark
(antiquark) numbers of {\it positive} and {\it negative} helicities.
$<L_z>_{q+\bar q}$ denotes the total orbital angular momentum carried 
by {\it quarks and antiquarks}, and $<J_z>_{G}$ is the gluon angular 
momentum. Further separation of $<J_z>_G$ into the spin and orbital 
pieces $\Delta G$ and $<L_z>_G$ is gauge dependent, and only in 
lightlike gauge and infinite momentum frame, $\Delta G$ could be 
identified as the gluon helicity measured in DIS processes. The 
smallness of ${1\over 2}\Delta\Sigma$ implies that the missing part 
should be contributed by either the quark orbital motion or gluon 
angular momentum. In the past decade, although considerable experimental 
and theoretical progress has been made in determining the quark spin 
contribution in the nucleon \cite{review}, one only obtains indirectly 
$\Delta G\simeq 0.5-1.5$ at $Q^2\simeq 10$ GeV$^2$ from the analysis 
of $Q^2$ dependence of $g_1(x,Q^2)$ \cite{smc97} with large errors. 
There is no direct data on $\Delta G$ except for a preliminary 
restriction on $\Delta G(x)/G(x)$ given by E581/704 experiment
\cite{e581}. Hopefully, several proposals of measuring the gluon 
helicity \cite{gluon} have been suggested. Most recently, it has 
been shown that $<J_z>_{q+\bar q}$ can be measured in the deep 
virtual compton scattering process \cite{ji-jws}, and one can obtain 
the quark orbital angular momentum from the difference $<J_z>-<s_z>$. 
Hence the experimental measurement and theoretical investigation of 
the quark orbital angular momentum are important and interesting.

Historically, when the quark model \cite{gellmann64} was invented 
in 1960's, all three quarks in the nucleon are assumed to be in 
S-states, so $<L_z>_q=0$ and the nucleon spin is completely 
attributed to the quark spin. On the other hand, in the naive parton 
model \cite{feynman69}, all quarks, antiquarks and gluons are moving 
in the same direction, i.e. parallel to the proton momentum, there is 
no transverse momentum for the partons and thus $<L_z>_{q+\bar q}=0$ 
and $<L_z>_G=0$. This picture cannot be $Q^2$ independent due to QCD 
evolution. In leading-log approximation, $\Delta\Sigma$ is $Q^2$
independent while the gluon helicity $\Delta G$ increases with $Q^2$. 
This increase should be compensated by the decrease of the orbital 
angular momentum carried by partons (see for instance earlier paper 
\cite{bms79} and later analysis \cite{rat87-sd89-qrrs90}). Recently,
the leading-log evolution of $<L_z>_{q+\bar q}$ and $<L_z>_G$, and an 
interesting asymptotic partition rule are obtained in \cite{jth96}.
The perturbative QCD can predict $Q^2$ dependence of the spin and 
orbital angular momenta, but not their values at the renormalization 
scale $\mu^2$, the spin structure of the nucleon is essentially 
determined by nonperturbative dynamics of the QCD bound state. 
The lattice QCD has provided a nonperturbative framework of 
evaluating the hadron structure and has obtained many interesting
results. Meantime, many QCD inspired nucleon models 
have been developed to explain existing data and yield good physical 
insight into the nucleon. For instance, in the bag model \cite{cmbag}, 
$<s_z>_{q}\simeq 0.39$, and $<L_z>_{q}\simeq 0.11$, while in the 
skyrme model \cite{skyrme,bali}, $\Delta G=\Delta\Sigma=0$, and 
$<L_z>={1\over 2}$, which implies that the nucleon spin arises 
only from the orbital motions. 

Phenomenologically, long before the EMC experimental data published 
\cite{emc}, using the Bjorken sum rule and low energy hyperon 
$\beta$-decay data (basically axial coupling constants), Sehgal
\cite{sehgal} shown that nearly $40\%$ of the nucleon spin arises 
from the orbital motion of quarks and rest $60\%$ is attributed to 
the spin of quarks and antiquarks. Most recently Casu and Sehgal 
\cite{cs97} shown that to fit the baryon magnetic moments and 
polarized DIS data, a large collective orbital angular momentum 
$<L_z>$, which contributes almost $80\%$ of nucleon spin, is needed. 
Hence the question of how much of the nucleon spin is coming from 
the quark orbital motion remains. This paper will discuss this 
question within the chiral quark model with symmetry breakings. In 
section II, the basic formalism of the chiral quark model in 
describing the quark spin and flavor contents is briefly reviewed 
and reorganized. A complete scheme for describing both spin and 
orbital angular momenta carried by quarks and antiquarks in the 
nucleon is developed in section III. The extension of this scheme to 
other octet and decuplet baryons is given in section IV. Possible 
modification and application of this scheme are discussed in section V.
\bigskip

\leftline{\bf II. Chiral quark model}

The chiral quark model was first formulated by Manohar and Georgi  
\cite{mg} and describes successfully the static properties of the
nucleon in the scale range between $\Lambda_{\rm QCD}$ ($\sim$ 
0.2-0.3 GeV) and $\Lambda_{\chi{\rm SB}}$ ($\sim$ 1 GeV). The 
relevant degrees of freedom are the constituent (dressed) quarks 
and Goldstone bosons associated with the spontaneous breaking of
the SU(3)$\times$SU(3) chiral symmetry. In this quasiparticle 
description, the effective gluon coupling is small and the dominant
interaction is coupling among quarks and Goldstone bosons.
This model was first employed by Eichten, Hinchliffe and Quigg 
\cite{ehq92} to explain both the {\it sea flavor asymmetry}
$\delta=\bar d-\bar u>0$ \cite{nmc} and the {\it smallness of}
$\Delta\Sigma$ in the nucleon. The model has been improved by 
introducing U(1)-breaking \cite{cl1} and kaonic suppression 
\cite{smw}. A complete description with both SU(3) and U(1)-breakings 
was developed in \cite{song9605} and \cite{cl2}, another 
$\lambda_8$-breaking version was given in \cite{wsk}. The description 
given in \cite{song9605} has been reformed into a compact one-parameter 
scheme in \cite{song9705} and the predictions are in good agreement 
with both spin and flavor observables. In the chiral quark model,
the effective Lagrangian describing interaction between quarks and 
the octet Goldstone bosons and singlet $\eta'$ is 
$${\it L}_I=g_8{\bar q}\pmatrix{({\rm GB})_+^0
& {\pi}^+ & {\sqrt\epsilon}K^+\cr 
{\pi}^-& ({\rm GB})_-^0
& {\sqrt\epsilon}K^0\cr
{\sqrt\epsilon}K^-& {\sqrt\epsilon}{\bar K}^0
&({\rm GB})_s^0 
\cr }q, 
\eqno (2a)$$
where $({\rm GB})_{\pm}^0$ and $({\rm GB})_{s}^0$ are defined as
$$({\rm GB})_{\pm}^0=\pm {\pi^0}/{\sqrt 2}+
{\sqrt{\epsilon_{\eta}}}{\eta^0}/{\sqrt 6}+
{\zeta'}{\eta'^0}/{\sqrt 3},
~~~({\rm GB})_s^0=-{\sqrt{\epsilon_{\eta}}}{\eta^0}/{\sqrt 6}+
{\zeta'}{\eta'^0}/{\sqrt 3}.
\eqno (2b)$$
The breaking effects are explicitly included. $a\equiv|g_8|^2$ denotes 
the transition probability of chiral fluctuation or splitting 
$u(d)\to d(u)+\pi^{+(-)}$, and $\epsilon a$ denotes the probability 
of $u(d)\to s+K^{-(0)}$. Similar definitions are used for 
$\epsilon_\eta a$ and $\zeta'^2a$. If the breaking parameter is
dominated by mass suppression effect, one reasonably expects
$0\leq\zeta'^2a<\epsilon_{\eta}a\simeq\epsilon a\leq a$, then 
we have $0\leq\zeta'^2\leq 1$, $0\leq\epsilon_{\eta}\leq 1$, and
$0\leq\epsilon\leq 1$. We note that in our formalism, only the 
{\it integrated} quark spin and flavor contents are discussed. The 
dependence of these contents on momentum variable or Bjorken $x$
in the chiral quark model has been discussed in \cite{ehq92,sbf96} 
and most recently in \cite{sw97}. 

The basic assumptions of the chiral quark model we used are: 
(i) the nucleon flavor, spin and orbital contents are determined 
by its valence quark structure and all possible chiral fluctuations 
$q\to q'+{\rm GB}$. The probabilities of these fluctuations are 
determined by the interaction Lagrangian (2), (ii) the coupling 
between the quarks and Goldstone bosons is rather weak, one can 
treat the fluctuation $q\to q'+{\rm GB}$ as a small perturbation 
($a\sim 0.10-0.15$) and the contributions from the higher order 
fluctuations can be neglected ($a^2<<1$), and (iii) the {\it 
valence quark structure} is assumed to be ${\rm SU}(3)_{flavor}
\otimes {\rm SU}(2)_{spin}$. Possible modifications of the third 
assumption will be discussed later.

The important feature of the chiral fluctuation is that due 
to the coupling between the quarks and GB's, a quark {\it flips} 
its spin and changes (or maintains) its flavor by emitting a 
charged (or neutral) Goldstone boson. The light quark sea asymmetry 
$\bar u<\bar d$ is attributed to the existing {\it flavor asymmetry of 
the valence quark numbers}, two valence $u$-quarks and one valence 
$d$-quark, in the proton. On the other hand, the quark spin reduction 
is due to the {\it spin dilution} in the chiral splitting processes 
$q_{\up}\to q_{\dw}$+GB. Most importantly, since the quark spin flips 
in the fluctuation with GB emission, hence the quark spin component
changes one unit of angular momentum, $(s_z)_f-(s_z)_i=+1$ or $-1$, 
the angular momentum conservation requires the {\it same amount change} 
of the orbital angular momentum but with {\it opposite sign}, i.e.
$(L_z)_f-(L_z)_i=-1$ or $+1$. This {\it induced orbital motion} 
distributes among the quarks and antiquarks, and compensates the
spin reduction in the chiral splitting, and restores the angular 
momentum conservation. This is the starting point to calculate the 
orbital angular momenta carried by quarks and antiquarks in the chiral 
quark model.

Before going to the discussion of the quark orbital motion, we 
briefly review the formalism developed in \cite{song9605,song9705}.
For a spin-up valence $u$-quark, the allowed fluctuations are
$$u_{\up}\to d_{\dw}+\pi^+,~~
u_{\up}\to s_{\dw}+K^+,~~
u_{\up}\to u_{\dw}+({\rm GB})_+^0,~~
u_{\up}\to u_{\up}.
\eqno (3)$$
Similarly, one can write down the allowed fluctuations for $u_{\dw}$,
$d_{\up}$, and $d_{\dw}$. Considering the valence quark {\it numbers} 
in the {\it proton} 
$$n_p(u_{\up})={5\over 3}~,~~~n_p(u_{\dw})={1\over 3}~,~~~
n_p(d_{\up})={1\over 3}~,~~~n_p(d_{\dw})={2\over 3}~.
\eqno (4)$$
the spin-up and spin-down quark (or antiquark) contents, up to first
order fluctuation, can be written as
$$n_p(q'_{\up,\dw}, {\rm or}\ {\bar q'}_{\up,\dw}) 
=\sum\limits_{q=u,d}\sum\limits_{h=\up,\dw}
n_p(q_h)P_{q_h}(q'_{\up,\dw}, {\rm or}\ {\bar q'}_{\up,\dw}),
\eqno (5)$$
where $P_{q_{\up,\dw}}(q'_{\up,\dw})$ and $P_{q_{\up,\dw}}({\bar
q}'_{\up,\dw})$ are the probabilities of finding a quark $q'_{\up,\dw}$
or an antiquark $\bar q'_{\up,\dw}$ arise from all chiral fluctuations 
of a valence quark $q_{\up,\dw}$. 
$P_{q_{\up,\dw}}(q'_{\up,\dw})$ and $P_{q_{\up,\dw}}({\bar q}'_{\up,\dw})$ 
can be obtained from the effective Lagrangian (2).
In Table I, only
$P_{q_{\up}}(q'_{\up,\dw})$ and $P_{q_{\up}}({\bar q}'_{\up,\dw})$ 
are listed. Those arise from $q_{\dw}$ can be obtained by using the
following relations
$$P_{q_{\dw}}(q'_{\up,\dw})=P_{q_{\up}}(q'_{\dw,\up}),~~~
P_{q_{\dw}}({\bar q}'_{\up,\dw})=P_{q_{\dw}}({\bar q}'_{\dw,\up})
\eqno (6)$$
where 
$$f\equiv{1\over 2}+{{\epsilon_{\eta}}\over 6}+{{\zeta'^2}\over 3},~~~
f_s\equiv{{2\epsilon_{\eta}}\over 3}+{{\zeta'^2}\over 3}
\eqno (7a)$$
and 
$$A\equiv 1-\zeta'+{{1-{\sqrt\epsilon_{\eta}}}\over 2},
\qquad  B\equiv \zeta'-{\sqrt\epsilon_{\eta}}\qquad
C\equiv \zeta'+2{\sqrt\epsilon_{\eta}}
\eqno (7b)$$
The special combinations $A$, $B$ and $C$ stem from the quark and
antiquark contents in the octet and singlet neutral bosons 
$({\rm GB})_{\pm}^0$ and $({\rm GB})_{s}^0$ (see (2b)) appeared in 
the effective chiral Lagrangian (2a), while $f$ and $f_s$ stand 
for the probabilities of the chiral splittings 
$u_{\up}(d_{\up})\to u_{\dw}(d_{\dw})+({\rm GB})^0_{+(-)}$ and
$s_{\up}\to s_{\dw}+({\rm GB})_s^0$ respectively. Although there is 
no valence $s$ quark in the proton and neutron, there are one or two 
valence $s$ quarks in $\Sigma$ or $\Xi$, or other strange decuplet
baryons, and even three valence $s$ quarks in the $\Omega^-$. Hence 
for the purpose of later use we also give the probabilities arise 
from a valence $s$-quark splitting. 

In general, the suppression effects may be different for different 
baryons, hence the probabilities $P_{q_{\up,\dw}}(q'_{\up,\dw})$ 
and $P_{q_{\up,\dw}}({\bar q}'_{\up,\dw})$ may vary with the baryons.
But we will assume that they are universal for all baryons. We also 
note that $P_{q_{\up,\dw}}(q'_{\up,\dw})$ and 
$P_{q_{\up,\dw}}({\bar q}'_{\up,\dw})$ satisfy the following
relations
$$P_{u_{\up}}(u_{\up,\dw})=P_{d_{\up}}(d_{\up,\dw}),~~
P_{u_{\up}}(d_{\up,\dw})=P_{d_{\up}}(u_{\up,\dw})
\eqno (8a)$$
$$P_{u_{\up}}({\bar u}_{\up,\dw})=P_{d_{\up}}({\bar d}_{\up,\dw}),~~
P_{d_{\up}}({\bar u}_{\up,\dw})=P_{u_{\up}}({\bar d}_{\up,\dw})
\eqno (8b)$$
$$P_{s_{\up}}(u_{\up,\dw})=P_{s_{\up}}(d_{\up,\dw})=
P_{u_{\up}}(s_{\up,\dw})=P_{d_{\up}}(s_{\up,\dw})
\eqno (8c)$$
$$P_{s_{\up}}({\bar u}_{\up,\dw})=P_{s_{\up}}({\bar d}_{\up,\dw})=
P_{u_{\up}}({\bar s}_{\up,\dw})=P_{d_{\up}}({\bar s}_{\up,\dw})
\eqno (8d)$$
In addition, it is easy to check the probabilities listed in Table I
satisfy
$$\sum\limits_{q'_h}P_{q_{\up}}(q'_h)-
\sum\limits_{\bar q'_h}P_{q_{\up}}(\bar q'_h)=1,~~~~~{\rm for}~q=u,d,s.
\eqno (8e)$$
The same holds for $P_{q_{\dw}}(q'_h)$ and $P_{q_{\dw}}(\bar q'_h)$.

Using (4), (5) and the probabilities listed in Table I, and defining
$$U_1={1\over 3}[A^2+2(3-A)^2],~~~~~
D_1={1\over 3}[2A^2+(3-A)^2]
\eqno (9a)$$
$$U_2=4D_2=4(\epsilon+2f-1)
\eqno (9b)$$
the spin-up and spin-down quark and antiquark contents, and the spin 
average and spin weighted quark and antiquark contents in the proton 
in the chiral quark model with both SU(3) and U(1)-breaking effects
were obtained in \cite{song9605,song9705} and are now collected in 
Table II. For the purpose of later discussion, we write down the quark
spin contents as 
$$\Delta u^p={4\over 5}\Delta_3-a,~~\Delta d^p=-{1\over 5}\Delta_3-a,~~
\Delta s^p=-\epsilon a,
\eqno (10a)$$
where $\Delta_3\equiv {5\over 3}[1-a(\epsilon+2f)]$. The total
spin content 
$${1\over 2}\Delta\Sigma^p={1\over 2}(\Delta u^p+\Delta d^p+\Delta
s^p)={1\over 2}-a(1+\epsilon+f)
\eqno (10b)$$
and the excess of down-sea over up-sea 
$$\delta\equiv \bar d-\bar u={{2A}\over 3}a.
\eqno (10c)$$
In the chiral quark model, all antiquark sea helicities are zero, 
$\Delta\bar q=0$ ($\bar q=\bar u,\bar d,\bar s$). 
\bigskip

\leftline{\bf III. Quark orbital motion in the nucleon.}

We now turn to the orbital angular momenta carried by quarks and 
antiquarks in the nucleon. The discussion of the orbital angular 
momentum contents is somewhat different from above. For instance, 
for a spin-up valence $u$-quark, only first three processes in (3), 
i.e. {\it quark fluctuations with {\rm GB} emission}, can induce 
change of the orbital angular momentum. The last process in (3), 
$u_{\up}\to u_{\up}$ means {\it no chiral fluctuation} and does not 
flip the quark spin. Hence it makes {\it no contribution} to the 
orbital motion and will be disregarded. We assume that the orbital 
angular momentum produced from the splitting $q_{\up}\to q'_{\dw}+
{\rm GB}$ is {\it equally shared} by all quarks and antiquarks, and 
introduce a {\it partition factor} $k$, which depends on the numbers 
of final state particles and interactions among them. If the Goldstone 
boson has a simple quark structure, i.e. each boson consists of a 
quark and an antiquark, one has two quarks and one antiquark, total 
number is {\it three}, after each splitting. Hence up to first order 
fluctuation, one has $k=1/3$, where the interactions between the
fluctuated quark and spectator quarks are neglected. We note that 
the `{\it equal sharing}' of the induced orbital angular momentum 
among splitting products is a crude approximation and needs to
be further improved. For simplicity, the equal sharing assumption 
will be used in this paper. 

We define $<L_z>_{q'/q_{\up}}$ ($<L_z>_{{\bar q'}/q_{\up}}$) 
as the orbital angular momentum carried by the quark $q'$ 
(antiquark $\bar q'$), arises from all fluctuations of a valence 
spin-up quark except for no emission case. Considering the quark 
spin component changes one unit of angular momentum in each 
splitting and using Table I, we can obtain all $<L_z>_{q'/q_{\up}}$ 
and $<L_z>_{\bar q'/q_{\up}}$ for $q=u,d,s$. They are listed in 
Table III. Again, for the purpose of later use, we also give the 
orbital angular momentum produced from a valence strange quark 
fluctuation.

Since the orbital angular momentum produced from a spin-up valence 
quark splitting is {\it positive}, while that from a spin-down 
valence quark splitting is {\it negative}, one has
$$ <L_z>_{q'/q_{\dw}}=-<L_z>_{q'/q_{\up}},~~ 
<L_z>_{{\bar q'}/q_{\dw}}=-<L_z>_{{\bar q'}/q_{\up}}
\eqno (11)$$
We note that {\it both $q'_{\up}$ and $q'_{\dw}$ are included}
in $<L_z>_{q'/q_{\up,\dw}}$ (the same for $<L_z>_{{\bar
q'}/q_{\up,\dw}}$), because the fractions of produced orbital 
angular momentum shared by the quarks (or antiquarks) do not 
depend on their spin states. 

Having obtained the orbital angular momenta carried by different 
quark flavors produced from the spin-up and spin-down valence quark
fluctuations, we can easily write down the total orbital angular 
momentum carried by a specific quark flavor, for instance $u$-quark 
in the proton
$$<L_z>_{u}^p=\sum\limits_{q=u,d}
[n_p(q_{\up})-n_p(q_{\dw})]<L_z>_{u/q_{\up}}
\eqno (12)$$
where $\sum$ summed over the $u$ and $d$ {\it valence quarks} in
the proton, $n_p(q_{\up})$ and $n_p(q_{\dw})$ are given 
in (4) for the simple SU(3)$\otimes$SU(2) proton wave function.
Note that different baryons will have different valence quark 
structure and thus different $n_B(q_{\up})$ and $n_B(q_{\dw})$.
Similarly, one can obtain the $<L_z>_{d}^p$, $<L_z>_{s}^p$, and 
corresponding quantities for the antiquarks. The numerical results
are listed in Table IV.

Defining $<L_z>_{q}^p$ ($<L_z>_{\bar q}^p$) as the total orbital 
angular momentum carried by all {\it quarks} ({\it antiquarks}), 
we finally obtain
$$<L_z>_{q}^p\equiv<L_z>_{u+d+s}^p=2ka(1+\epsilon+f)
\eqno (13a)$$
$$<L_z>_{\bar q}^p\equiv
<L_z>_{\bar u+\bar d+\bar s}^p=ka(1+\epsilon+f)
\eqno (13b)$$
$$<L_z>_{q+\bar q}^p\equiv<L_z>_{q}^p+<L_z>_{\bar q}^p=3ka(1+\epsilon+f)
\eqno (13c)$$
The sum of (13c) and (10b) gives
$$<J_z>_{q+\bar q}^p={1\over 2}-a(1-3k)(1+\epsilon+f)
\eqno (14)$$
Taking $k=1/3$, we obtain $<J_z>_{q+\bar q}^p=1/2$. This result means 
that in the chiral fluctuations, the missing part of the quark spin 
{\it is transferred} into the orbital motion of quarks and antiquarks. 
The amount of quark spin reduction $a(1+\epsilon+f)$ in (10b) is 
canceled by the same amount increase of the quark orbital angular 
momentum in (13c), and the total angular momentum of nucleon is 
unchanged. We note that this result is a consistency check for the 
formalism and not a logical deduction.

Two remarks should be made here. Although the orbital angular momentum
carried by quarks (or antiquarks) $<L_z>_{q}^p$ (or $<L_z>_{\bar q}^p$) 
depends on the the chiral parameters, the ratio 
$<L_z>_{q}^p/<L_z>_{\bar q}^p=2$ is {\it independent of the probabilities 
of chiral fluctuations}. This is originated from the mechanism of the 
chiral fluctuation: there are {\it two} quarks and {\it one} antiquark 
in the final state, and they equally share the orbital angular momentum 
produced in the splitting process. Second, the total {\it loss} of 
quark spin $a(1+\epsilon+f)$ appeared in (10b) is due to the fact that 
there are {three} splitting processes (for instance see (3)), which 
flip the quark spin, the probabilities of these fluctuations are $a$, 
$\epsilon a$, and $fa$ respectively. For the same reason, the total 
{\it gain} of the orbital angular momentum is $3ka(1+\epsilon+f)$.

The above results can be easily extended to the neutron. Explicit 
calculation gives
$$<L_z>_{u,\bar u}^n=<L_z>_{d,\bar d}^p,~~
<L_z>_{d,\bar d}^n=<L_z>_{u,\bar u}^p,~~
<L_z>_{s,\bar s}^n=<L_z>_{s,\bar s}^p
\eqno (15)$$
Using these relations, one can obtain the orbital angular
momenta carried by quarks and antiquarks in the neutron.
Since we have similar relations for $\Delta q$ from isospin symmetry,
$$\Delta u^n=\Delta d^p,~~\Delta d^n=\Delta u^p,~~
\Delta s^n=\Delta s^p
\eqno (16)$$
hence the main results (10b), (13a-c), (14), and related conclusions 
hold for the neutron as well. The new scheme for describing both quark
spin and orbital contents in the nucleon derived in this section can be 
easily extended to other octet and decuplet baryons, which will be 
given in section IV.

To see how much of the nucleon spin is contributed by the quark
(or antiquark) orbital motions, let us estimate (13a-c).
From (7a), one obtains
$$1+\epsilon+f=3.0,~~~~~~~ {\rm for}~~{\rm U(3)-symmetry}~
(\epsilon=\epsilon_{\eta}=\zeta'^2=1)
\eqno (17a)$$
$$1+\epsilon+f=1.5,~~~~~~~ {\rm for}~~{\rm Extreme~ breaking}~
(\epsilon=\epsilon_{\eta}=\zeta'^2=0)
\eqno (17b)$$
The reality might be in between. The NMC data gives 
$\bar d-\bar u=0.147\pm 0.039$ \cite{nmc} with large uncertainty.
Taking $\bar d-\bar u=0.130\pm 0.020$ (see Notes added on p.12), 
$\Delta s=-0.06$, and $a=0.12$ which gives
$$ 1+\epsilon+f=2.16\pm 0.08.
\eqno (18)$$
From (13a) and (13b), one obtains 
$$<L_z>_{q}^p=0.172\pm 0.006,~~~~~~<L_z>_{\bar q}^p=0.086\pm 0.003,
~~~~~~<L_z>_{q}^p/<L_z>_{\bar q}^p=2.
\eqno (19)$$
where the last equality holds exactly in the chiral quark model.
Considering $<J_z>_{q+\bar q}^p=0.5$, one has
$<L_z>_{q+\bar q}^p=0.26\pm 0.01$, and $<s_z>_{q+\bar q}^p=0.24\pm 0.01$.
The orbital angular momenta shared by different quark flavors are 
listed in Table IV. Comparison with other models is also given. 
However, the orbital angular momenta $<L_z>_{q(\bar q)}$ and total 
angular momentum $<L_z>_{q+\bar q}$ listed in Table IV depends on 
the chiral parameters we used. The numerical results show that
$$<L_z>_{q+\bar q}^p=0.27\pm 0.04,~~~~<s_z>_{q+\bar q}^p=0.23\pm 0.04,
\eqno (20a)$$
and
$$<s_z>_{q+\bar q}^p/<L_z>_{q+\bar q}^p=0.90\pm 0.27.
\eqno (20b)$$
i.e. nearly $55\%$ of the proton spin is coming from the orbital 
motion of quarks and antiquarks, and $45\%$ is contributed by 
the quark spin, with large theoretical uncertainty. Our result
can be compared with the result given in \cite{sehgal}. 

\bigskip

\leftline{\bf IV. Spin and orbital contents in other baryons.}

\leftline{\bf A. Octet baryons:}

We take $\Sigma^+(uus)$ as an example, other octet baryons 
can be discussed in a similar manner. The valence quark structure 
of $\Sigma^+$ is the same as the proton with the replacement
$d\to s$. Hence one has
$$n_{\Sigma^+}(u_{\up})={5\over 3}~,~~~n_{\Sigma^+}(u_{\dw})={1\over 3}~,~~~
n_{\Sigma^+}(s_{\up})={1\over 3}~,~~~n_{\Sigma^+}(s_{\dw})={2\over 3}~.
\eqno (21)$$
The spin-up and spin-down quark (antiquark) contents in the 
$\Sigma^+$ are
$$n_{\Sigma^+}(q'_{\up,\dw}, {\rm or}\ {\bar q'}_{\up,\dw}) 
=\sum\limits_{q=u,s}\sum\limits_{h=\up,\dw}
n_{\Sigma^+}(q_h)P_{q_h}(q'_{\up,\dw}, {\rm or}\ {\bar q'}_{\up,\dw})
\eqno (22)$$
The spin-weighted quark contents in $\Sigma^+$ are 
$$\Delta u^{\Sigma^+}={4\over 3}-{a\over 3}(4+3\epsilon+8f),~~
\Delta d^{\Sigma^+}=-{a\over 3}(4-\epsilon),~~
\Delta s^{\Sigma^+}=-{1\over 3}-{{2a}\over 3}(\epsilon-f_s)
\eqno (23)$$
and all antiquark sea helicities are zero. Hence
$$<s_z>_{q+\bar q}^{\Sigma^+}={1\over 2}(\Delta\Sigma)^{\Sigma^+}
={1\over 2}-{{a}\over 3}[4(1+\epsilon+f)-(2\epsilon+f_s)]
\eqno (24)$$
From now on we define 
$$\xi_1\equiv 1+\epsilon+f,~~~~~~~~~~\xi_2\equiv 2\epsilon+f_s
\eqno (25)$$
then relation (10b) can be rewritten as
$$(\Delta\Sigma)^{p}=1-{{2a}\over 3}(3\xi_1),
\eqno (26a)$$
and (24) leads to
$$(\Delta\Sigma)^{\Sigma^+}=1-{{2a}\over 3}(4\xi_1-\xi_2)
\eqno (26b)$$
Similarly, we can obtain the spin-weighted quark contents and the
total spin contents of quark and antiquarks in other members of
the baryon octet. The results of $\Sigma^+$, $\Sigma^0$, $\Lambda^0$,
and $\Xi^0$ are listed in Table V. (We note that the quark spin 
contents, but not orbital contents, in the octet baryons were discussed 
in \cite{los1,wb}). Those for $\Sigma^-$, and $\Xi^-$, can be obtained 
by using the following relations due to isospin symmetry 
$$\Delta u^{\Sigma^-}=\Delta d^{\Sigma^+},~~~
\Delta d^{\Sigma^-}=\Delta u^{\Sigma^+},~~~
\Delta s^{\Sigma^-}=\Delta s^{\Sigma^+},~~~
\eqno (27a)$$
$$\Delta u^{\Xi^-}=\Delta d^{\Xi^0},~~~
\Delta d^{\Xi^-}=\Delta u^{\Xi^0},~~~
\Delta s^{\Xi^-}=\Delta s^{\Xi^0},~~~
\eqno (27b)$$
which can be verified by explicit calculations. Furthermore, in the 
SU(3) symmetry case, $\epsilon=\epsilon_{\eta}=1$ and $f=f_s$, one 
obtains, from Table V,
$$\Delta u^{\Sigma^+}=\Delta u^{p},~~~
\Delta d^{\Sigma^+}=\Delta s^{p},~~~
\Delta s^{\Sigma^+}=\Delta d^{p},
\eqno (28a)$$
$$\Delta u^{\Xi^0}=\Delta d^{p},~~~
\Delta d^{\Xi^0}=\Delta s^{p},~~~
\Delta s^{\Xi^0}=\Delta u^{p}.
\eqno (28b)$$
We note that these relations are consequences of SU(3) symmetry and 
{\it do not depend on} the U(1)-breaking parameter $\zeta'$.

The total spin contents of quarks and antiquarks in the octet baryons
can be written as (see Table V)
$$(\Delta\Sigma)^{B}=1-{{2a}\over 3}(c_1\xi_1+c_2\xi_2)
\eqno (29)$$
which is generalization of (26a) and (26b). In (29), the coefficients 
$c_1$ and $c_2$ satisfy $c_1+c_2=3$, and ($c_1$, $c_2$)=(3, 0), 
(4, $-1$), (0, 3), and ($-1$, 4) for $B=$N, $\Sigma$, $\Lambda$, and 
$\Xi$ respectively. Hence {\it the spin reductions for all members in 
the same isospin multiplet are the same}, but may be different for
different isospin multiplets, except for the cases of SU(3)-symmetry 
limit ($\xi_1=\xi_2=2+f$), or U(3)-symmetry limit ($\xi_1=\xi_2=3$). 
For example, in the U(3)-symmetry limit,
$$(\Delta\Sigma)^{N}=(\Delta\Sigma)^{\Sigma}
=(\Delta\Sigma)^{\Lambda}=(\Delta\Sigma)^{\Xi}=1-6a
\eqno (30)$$
On the other hand, in the extreme breaking case, $\xi_1=1.5$, 
$\xi_2=0$, one obtains
$$(\Delta\Sigma)^{N}=1-3a,~~~(\Delta\Sigma)^{\Sigma}=1-4a,~~~
(\Delta\Sigma)^{\Lambda}=1,~~~(\Delta\Sigma)^{\Xi}=1+a.
\eqno (31)$$
For the real world, the results might be in between (30) and (31).
It is interesting to note that in the extreme breaking case (31), 
there is no quark spin reduction in $\Lambda^0$, and even a small 
increase ($a/2$) of the total quark spin content in $\Xi$. 

Similar to the nucleon case, the orbital angular momenta carried 
by quarks and antiquarks in other octet baryons can be calculated.
The results for different isospin multiplets are listed in Table VI,
and the total orbital angular momentum carried by all
quarks and antiquarks in the baryon $B$ is
$$<L_z>_{q+\bar q}^{B}=ka(c_1\xi_1+c_2\xi_2)
\eqno (32)$$
The sum of both spin and orbital angular momenta (29) and (32) gives
$$<J_Z>_{q+\bar q}^{B}=<s_z>_{q+\bar q}^{B}+
<L_z>_{q+\bar q}^{B}={1\over 2}-{{a}\over 3}(1-3k)(c_1\xi_1+c_2\xi_2)
\eqno (33)$$
Again, we obtain $<J_z>_{q+\bar q}^{B}=1/2$ for all octet baryons, if
$k=1/3$. i.e. the loss of the quark spin is compensated by the
gain of the orbital motion of quarks and antiquarks. In addition,
one has $<L_z>_q^B/<L_z>_{\bar q}^B=2$. All results and
conclusions obtained in section III for the nucleon hold for other 
octet baryons as well. Similar to (15), the explicit calculation 
also gives the following isospin symmetry relations for the orbital
angular momenta in $\Sigma$ and $\Xi$ multiplets
$$<L_z>_u^{\Sigma^-}=<L_z>_d^{\Sigma^+},~~~
<L_z>_d^{\Sigma^-}=<L_z>_u^{\Sigma^+},~~~
<L_z>_s^{\Sigma^-}=<L_z>_s^{\Sigma^+}
\eqno (34a)$$
$$<L_z>_u^{\Xi^-}=<L_z>_d^{\Xi^0},~~~
<L_z>_d^{\Xi^-}=<L_z>_u^{\Xi^0},~~~
<L_z>_s^{\Xi^-}=<L_z>_s^{\Xi^0}
\eqno (34b)$$

From Table V, one has
$$\Delta u^B-\Delta d^B=c_B[1-(\epsilon+2f)]
\eqno (35)$$
where $c_B=5/3$, $4/3$, and $-1/3$ for $B=p$, $\Sigma^+$, and
$\Xi^0$ respectively. Using the isospin symmetry relations (15),
and (27a-b), one obtains from (35) the following identity
$$\Delta u^p-\Delta u^n+\Delta u^{\Sigma^-}-\Delta u^{\Sigma^+}+
\Delta u^{\Xi^0}-\Delta u^{\Xi^-}=0
\eqno (36)$$
This relation holds for $d-$quark spin and $s-$quark spin contents
as well. 

Similarly, by explicit calculation, one can show that the orbital 
angular momentum $<L_z>_u^B$ in the octet baryons satisfy similar
identity
$$<L_z>_u^p-<L_z>_u^n+<L_z>_u^{\Sigma^-}-<L_z>_u^{\Sigma^+}+
<L_z>_u^{\Xi^0}-<L_z>_u^{\Xi^-}=0
\eqno (37)$$
The same relations hold for $<L_z>_{d}^B$ and $<L_z>_{s}^B$, and
corresponding quantities for antiquarks. From (36) and (37), 
the magnetic moments (see eq.(48) below) of the octet baryons 
satisfy the well-known Coleman-Glashow sum rule
$$\mu_p-\mu_n+\mu_{\Sigma^-}-\mu_{\Sigma^+}+
\mu_{\Xi^0}-\mu_{\Xi^-}=0
\eqno (38)$$
We note that this sum rule was discussed in \cite{los1} without
the orbital contributions. Our result shows that (38) still holds 
for the symmetry breaking chiral quark model even the orbital 
contributions are included. 

\smallskip

\leftline{\bf B. Decuplet baryons:}

The above discussion can also be extended to the baryon decuplet. 
The quark spin and orbital angular momenta are listed in Table
VII and Table VIII respectively (we note that the spin contents, 
but not orbital contents, in the decuplet baryons were discussed 
in \cite{los2}). Again, the explicit calculation gives
$$(\Delta u)^{\Delta^{-}}=(\Delta d)^{\Delta^{++}},~~~
(\Delta d)^{\Delta^{-}}=(\Delta u)^{\Delta^{++}},~~~
(\Delta s)^{\Delta^{-}}=(\Delta s)^{\Delta^{++}}
\eqno (39a)$$
$$(\Delta u)^{\Delta^{0}}=(\Delta d)^{\Delta^{+}},~~~
(\Delta d)^{\Delta^{0}}=(\Delta u)^{\Delta^{+}},~~~
(\Delta s)^{\Delta^{0}}=(\Delta s)^{\Delta^{+}}
\eqno (39b)$$ 
for the $\Delta$ multiplet, 
$$(\Delta u)^{\Sigma^{*-}}=(\Delta d)^{\Sigma^{*+}},~~~
(\Delta d)^{\Sigma^{*-}}=(\Delta u)^{\Sigma^{*+}},~~~
(\Delta s)^{\Sigma^{*-}}=(\Delta s)^{\Sigma^{*+}}
\eqno (40)$$ 
for $\Sigma^*$ multiplet, and 
$$(\Delta u)^{\Xi^{*-}}=(\Delta d)^{\Xi^{*0}},~~~
(\Delta d)^{\Xi^{*-}}=(\Delta u)^{\Xi^{*0}},~~~
(\Delta s)^{\Xi^{*-}}=(\Delta s)^{\Xi^{*0}}
\eqno (41)$$ 
for $\Xi^*$ multiplet. These relations are originated from the 
isospin symmetry of the baryon wave functions. Hence, in Table 
VII, we only list the results for $\Delta^{++}$, $\Delta^{+}$, 
$\Sigma^{*+}$, $\Sigma^{*0}$, $\Xi^{*0}$, and $\Omega^{-}$. It is 
interesting to see that there is an {\it equal spacing rule} for 
total quark spin contents in the decuplet baryons
$$(\Delta\Sigma)^{B^*}=3-2a[3\xi_1+S(\xi_1-\xi_2)]
\eqno (42)$$
where the $S$ is the {\it strangeness} quantum number of the 
decuplet baryon $B^*$. The same rule was also obtained in \cite{los2}. 
Hence we have
$$(\Delta\Sigma)^{\Omega-\Xi}=
(\Delta\Sigma)^{\Xi-\Sigma}=
(\Delta\Sigma)^{\Sigma-\Delta}=2a(\xi_1-\xi_2).
\eqno (43)$$
For the $\Delta$ multiplet, $S=0$, (42) gives
$$(\Delta\Sigma)^{\Delta}=3[1-{{2a}\over 3}(3\xi_1)]
\eqno (44)$$
Comparing (44) with (29) (taking $c_1=3$, and $c_2=0$ in (29) for the
nucleon), one has $(\Delta\Sigma)^{\Delta}=3(\Delta\Sigma)^N$, i.e. 
total spin content of $\Delta$ baryon is {\it three} times that of 
the nucleon, which is physically reasonable result.

For the orbital angular momentum (see Table VIII), we have
$$<L_z>_{q+\bar q}^{B^*}=3ka[3\xi_1+S(\xi_1-\xi_2)]
\eqno (45)$$
Hence we have a similar {\it equal spacing rule} for the orbital
angular momenta
$$<L_z>_{q+\bar q}^{\Omega-\Xi}=
<L_z>_{q+\bar q}^{\Xi-\Sigma}=
<L_z>_{q+\bar q}^{\Sigma-\Delta}=-3ka(\xi_1-\xi_2).
\eqno (46)$$
The sum of spin and orbital angular momenta (42) and (45) gives
$$<J_z>_{q+\bar q}^{B^*}={3\over 2}-a(1-3k)[3\xi_1+S(\xi_1-\xi_2)]
\eqno (47)$$
Once again, the spin reduction is compensated by the increase of
orbital angular momentum and keep the total angular momentum 
of the baryon (now is 3/2 for the decuplet) unchanged. 
\bigskip 

\leftline{\bf V. Discussion and summary}

(1) We have assumed that there are {\it no gluons} and other 
degrees of freedom in the proton, hence $<J_z>_G$=0. This is 
presumably a good approximation at very low $Q^2$. However, if 
$<J_z>_G$ is {\it nonzero} \cite{bs90} and {\it not small}, the 
results given above should be modified. There are many ways to 
construct a bound state consists of three valence quarks and 
single- or multi-gluon. We do not attempt to present a detail model
calculation here, but only make a crude estimation. Assuming that 
the gluons are pure {\it spectators}, the only role of them is to 
provide a nonzero and positive $<J_z>_G$. Taking $<J_z>_G(1~{GeV^2})
\simeq 0.20\pm 0.10$ given in \cite{bj97}, and assuming the ratios 
(20b) derived from the chiral quark model hold approximately, 
one obtains $<L_z>_{q+\bar q}(1~{GeV^2})\simeq 0.16\pm 0.02$, and
$<s_z>_{q+\bar q}(1~{GeV^2})\simeq 0.14\pm 0.02$. The total quark 
and antiquark spin content can be compared with DIS data
\cite{smc-slac-hermes,smc97}, and lattice QCD result \cite{dll95}. 
However, if the gluonic degrees of freedom is included, then gluons 
should also share the induced orbital angular momentum and the 
partition among splitting products would be different from what 
we assumed in this paper. This issue will be discussed elsewhere
\cite{song9803}. 

(2) One of important applications of our description is to study the 
baryon magnetic moments, which should depend on {\it both spin and
orbital} motions of quarks and antiquarks. The baryon magnetic 
moments can be written as 
$$\mu^{B(B^*)}=\sum\limits_{q=u,d,s}\mu_q[(\Delta q)^{B(B^*)}
+<L_z>^{B(B^*)}_{q}-<L_z>^{B(B^*)}_{\bar q}]
\eqno (48)$$
where $\Delta\bar q=0$ for $q=u,d,s$ have been used and $\mu_q$s are 
the magnetic moments of quarks. Using the spin and orbital contents 
given in Table IV, we obtain, from (48), the ratio of the proton to 
neutron magnetic moments ${\mu_p}/{\mu_n}=-1.45$ for $k=1/3$, while 
${\mu_p}/{\mu_n}=-1.43$ for $k=0$, where $\mu_s/\mu_d=2/3$ and 
$\mu_d=-0.45\mu_u$ are used. Since $({\mu_p}/{\mu_n})_{exp.}=-1.46$,
the agreement with data is improved by including the orbital 
contributions. A detail discussion of the baryon magnetic moments 
will be presented in another paper \cite{song9803}.

(3) In section III, for simplicity we have assumed that the orbital
angular momentum produced from the splitting $q_{\up}\to q'_{\dw}+
{\rm GB}$ is {\it equally shared} by all quarks and antiquarks. 
Another possible partition pattern is that the induced orbital 
angular momentum in the chiral splitting is shared by the recoil 
quark and Goldstone boson, then the quark and antiquark in the GB 
share the orbital angular momentum carried by the GB. In this case
the fractions of the orbital angular momentum carried by the quarks
and antiquark in each splitting process are not necessarily equal. 
This would make the analytic results given in Tables III, VI,
and VIII more involved, and change the numerical results of 
$<L_z>_{u,d,s}$ and $<L_z>_{\bar u,\bar d,\bar s}$ given in Table IV. 
However, it would not change $<L_z>_{q+\bar q}$ in Table IV and the
analytic results given in Tables I, II, V, VII, and spin contents in 
Table IV. 

To summary, we have developed a new and unified scheme for describing 
both spin and orbital motions of quarks and antiquarks in the nucleon
in symmetry breaking chiral quark model. The extension to the octet 
and decuplet baryons are presented. The results might provide some new
insight into the structure of the nucleon and other baryons.  
\bigskip

\leftline{\bf Acknowledgments}

The author thanks Xiangdong Ji and H. J. Weber for useful comments, and 
thank J. S. McCarthy for reading the manuscript. This work was supported 
in part by the U.S. DOE Grant No. DE-FG02-96ER-40950, the Institute of 
Nuclear and Particle Physics, University of Virginia, and the Commonwealth 
of Virginia.
\bigskip

{\it Notes added. $-$~} After this paper was submitted for publication, 
the author learned that (1) similar feature of the orbital angular
momentum in a simple SU(3) symmetry case was independently discussed 
in \cite{cl3}, (2) most recent E866 data gives $\bar d-\bar u=0.100\pm
0.018$ \cite{e866}, our input $\bar d-\bar u\simeq 0.130$ used in (18) 
is just in between the NMC and E866 values.

\begin{table}[ht]
\begin{center}
\caption{The probabilities $P_{q_{\up}}(q'_{\up,\dw},\bar q'_{\up,\dw})$
and $P_{q_{\up}}(q'_{\up,\dw},{\bar q}'_{\up,\dw})$} 
\begin{tabular}{cccc} 
$q'$ &$P_{u_{\up}}(q'_{\up,\dw})$ & $P_{d_{\up}}(q'_{\up,\dw})$ &
$P_{s_{\up}}(q'_{\up,\dw})$ \\ 
\hline 
$u_{\up}$ & $1-({{1+\epsilon}\over 2}+f)a+
{a\over {18}}(3-A)^2$ & ${a\over {18}}A^2$ &${a\over {18}}B^2$ \\
$u_{\dw}$ & $({{1+\epsilon}\over 2}+f)a+
{a\over {18}}(3-A)^2$ & $a+{a\over {18}}A^2$ &$\epsilon a+{a\over
{18}}B^2$ \\
$d_{\up}$ & ${a\over {18}}A^2$ &$1-({{1+\epsilon}\over 2}+f)a+
 {a\over {18}}(3-A)^2$ &${a\over {18}}B^2$ \\
$d_{\dw}$ & $a+{a\over {18}}A^2$ &
$({{1+\epsilon}\over 2}+f)a+{a\over {18}}(3-A)^2$ & 
$\epsilon a+{a\over {18}}B^2$ \\
$s_{\up}$ & ${a\over {18}}B^2$ &${a\over {18}}B^2$ &
$1-(\epsilon+f_s)a+{a\over {18}}C^2$ \\
$s_{\dw}$ & $\epsilon a+{a\over {18}}B^2$ & $\epsilon a+{a\over {18}}B^2$ 
&$(\epsilon+f_s)a+{a\over {18}}C^2$ \\
\hline
${\bar u}_{\up,\dw}$ &${a\over {18}}(3-A)^2$ & ${a\over 2}+{a\over
{18}}A^2$ &${{\epsilon a}\over 2}+{a\over {18}}B^2$ \\
${\bar d}_{\up,\dw}$ &${a\over 2}+{a\over {18}}A^2$&
${a\over {18}}(3-A)^2$ &${{\epsilon a}\over 2}+{a\over {18}}B^2$ \\
${\bar s}_{\up,\dw}$ &${{\epsilon a}\over 2}+{a\over {18}}B^2$&
${{\epsilon a}\over 2}+{a\over {18}}B^2$&
${a\over {18}}C^2$ \\
\end{tabular}
\end{center}
\end{table}

\begin{table}[ht]
\begin{center}
\caption{The spin-up and spin-down quark (antiquark) 
contents ${q}_{\up,\dw}$ and ${\bar q}_{\up,\dw}$, and 
spin-average and spin-weighted quark (antiquark) contents in the proton.
Where $q=q_{\up}+q_{\dw}$, $\bar q=\bar q_{\up}+\bar q_{\dw}$, 
$\Delta q=q_{\up}-q_{\dw}$ and 
$\Delta{\bar q}={\bar q}_{\up}-{\bar q}_{\dw}$ are used}
\begin{tabular}{ccc} 
$u_{\up}={5\over 3}+{a\over 3}(-2+{{U_1}\over 2}-{{U_2}\over 2})$& 
$d_{\up}={1\over 3}+{a\over 3}(2+{{D_1}\over 2}+{{D_2}\over 2})$& 
$s_{\up}=\epsilon a+{a\over 3}({{B^2}\over 2})$\\
$u_{\dw}={1\over 3}+{a\over 3}(5+{{U_1}\over 2}+{{U_2}\over 2})$& 
$d_{\dw}={2\over 3}+{a\over 3}(4+{{D_1}\over 2}-{{D_2}\over 2})$& 
$s_{\dw}=2\epsilon a+{a\over 3}({{B^2}\over 2})$\\
\hline 
${\bar u}_{\up}={\bar u}_{\dw}={a\over 2}+{a\over 3}({{U_1}\over 2})$&
${\bar d}_{\up}={\bar d}_{\dw}=a+{a\over 3}({{D_1}\over 2})$ &
${\bar s}_{\up}={\bar s}_{\dw}={{3\epsilon a}\over 2}+{a\over
3}({{B^2}\over 2})$\\
\hline
\hline
$u=2+{a\over 3}(3+U_1)$&$d=1+{a\over 3}(6+D_1)$& $s=3\epsilon a+{a\over
3}B^2$\\
\hline
$\bar u={a\over 3}(3+U_1)$&$\bar d={a\over 3}(6+D_1)$& $\bar s=3\epsilon
a+{a\over 3}B^2$\\
\hline
\hline
$\Delta u={4\over 3}[1-a(\epsilon+2f)]-a$ &$\Delta d={{-1}\over
3}[1-a(\epsilon+2f)]-a$ & $\Delta s=a(1-\epsilon)-a$ \\
\hline
$\Delta{\bar u}=0$ & $\Delta{\bar d}=0$ &$\Delta{\bar s}=0$ \\
\end{tabular}
\end{center}
\end{table}

\begin{table}[ht]
\begin{center}
\caption{The orbital angular momentum carried by the quark $q'$ 
($\bar q'$), {\it both spin-up and spin-down are included}, 
from a valence spin-up quark $q_{\up}$ fluctuates into all 
allowed final states.}
\begin{tabular}{cccc} 
&$<L_z>_{q',{\bar q'}/u_{\up}}$ &$<L_z>_{q',{\bar q'}/d_{\up}}$ 
&$<L_z>_{q',{\bar q'}/s_{\up}}$\\ 
\hline 
$q'=u$ & $ka[1+\epsilon+f+{{(3-A)^2}\over 9}]$ & $ka[1+{{A^2}\over 9}]$ 
&$ka[\epsilon+{{B^2}\over {9}}]$ \\
$q'=d$ & $ka[1+{{A^2}\over {9}}]$ &$ka[1+\epsilon+f+{{(3-A)^2}\over {9}}]$
&$ka[\epsilon+{{B^2}\over {9}}]$ \\
$q'=s$ &$ka[\epsilon+{{B^2}\over {9}}]$ 
&$ka[\epsilon+{{B^2}\over {9}}]$ &
$ka[2\epsilon+f_s+{{C^2}\over {9}}]$\\
\hline
${\bar q'}={\bar u}$ &$ka[{{(3-A)^2}\over {9}}]$&$ka[1+{{A^2}\over {9}}]$ 
&$ka[\epsilon+{{B^2}\over {9}}]$ \\
${\bar q'}={\bar d}$ &$ka[1+{{A^2}\over {9}}]$&$ka[{{(3-A)^2}\over {9}}]$
&$ka[\epsilon+{{B^2}\over {9}}]$ \\
${\bar q'}={\bar s}$&$ka[\epsilon+{{B^2}\over {9}}]$ 
&$ka[\epsilon+{{B^2}\over {9}}]$ 
&$ka[{{C^2}\over {9}}]$ \\
\end{tabular}
\end{center}
\end{table}

\begin{table}[ht]
\begin{center}
\caption{Quark spin and orbital angular momenta in the proton in the
chiral quark model and other models.}
\begin{tabular}{cccccc} 
Quantity & Data \cite{smc97}   & This paper &  Sehgal \cite{sehgal} 
&CS \cite{cs97}& NQM\\ 
\hline 
$<L_z>_u^p$       & $-$ & 0.110   &0.237   & $-$    & 0  \\ 
$<L_z>_{\bar u}^p$& $-$ & $-$0.006&0       & $-$    & 0  \\ 
$<L_z>_d^p$       & $-$ & 0.036   &$-$0.026& $-$    & 0  \\ 
$<L_z>_{\bar d}^p$& $-$ & 0.066   &0       & $-$    & 0  \\ 
$<L_z>_s^p$       & $-$ & 0.026   &0       & $-$    & 0  \\ 
$<L_z>_{\bar s}^p$& $-$ & 0.026   &0       & $-$    & 0  \\ 
\hline
$<L_z>_{q+\bar q}^p$ & $-$ & 0.26 &0.21  & 0.39    & 0  \\ 
\hline
$\Delta u^p$ & $0.85\pm 0.05$ & 0.91   & 0.91 & 0.78    & 4/3  \\ 
$\Delta d^p$ & $-0.41\pm 0.05$ & $-0.38$   & $-0.34$ & $-0.48$ & $-1/3$\\ 
$\Delta s^p$ & $-0.07\pm 0.05$ & $-0.06$   & 0 & $-0.14$ & 0  \\ 
\hline
${1\over 2}\Delta\Sigma^p$ & $0.19\pm 0.06$ & 0.24 & 0.29 & 0.08 & 1/2\\ 
\end{tabular}
\end{center}
\end{table}

\begin{table}[ht]
\begin{center}
\caption{The spin-weighted quark contents in the octet baryons, where
$\xi_1=1+\epsilon+f$ and $\xi_2=2\epsilon+f_s$.}
\begin{tabular}{ccccc} 
 Baryon &$\Delta u^B$ & $\Delta d^B$& $\Delta s^B$& $\Delta\Sigma^B$ \\
\hline 
p & ${4\over 3}-{a\over 3}(3+4\epsilon+8f)$& 
$-{1\over 3}-{a\over 3}(3-\epsilon-2f)$ & $-a\epsilon$&
$1-2a\xi_1$\\
$\Sigma^+$ & ${4\over 3}-{a\over 3}(4+3\epsilon+8f)$& 
$-{a\over 3}(4-\epsilon)$ & $-{1\over 3}-{{2a}\over 3}(\epsilon-f_s)$&
$1-{{2a}\over 3}(4\xi_1-\xi_2)$\\
$\Sigma^0$ & ${2\over 3}-{a\over 3}(4+\epsilon+4f)$& 
${2\over 3}-{a\over 3}(4+\epsilon+4f)$ 
& $-{1\over 3}-{{2a}\over 3}(\epsilon-f_s)$&
$1-{{2a}\over 3}(4\xi_1-\xi_2)$\\
$\Lambda^0$ & $-a\epsilon$&  $-a\epsilon$& 
$1-2a(\epsilon+f_s)$& $1-2a\xi_2$\\
$\Xi^0$ & $-{1\over 3}-{a\over 3}(-1+3\epsilon-2f)$& 
$-{a\over 3}(4\epsilon-1)$ & ${4\over 3}-{a\over 3}(7\epsilon+8f_s)$&
$1-{{2a}\over 3}(4\xi_2-\xi_1)$\\
\end{tabular}
\end{center}
\end{table}

\begin{table}[ht]
\begin{center}
\caption{The 
orbital angular momenta carried by quarks and antiquarks in
the octet baryons, where $\xi_1=1+\epsilon+f$ and $\xi_2=2\epsilon+f_s$.}
\begin{tabular}{ccccc} 
 Baryon &$<L_z>_q^B$ & $<L_z>_{\bar q}^B$&$<L_z>_{q+\bar q}^B$& 
$<J_z>_{q+\bar q}^B$ \\
\hline 
p & $2ka\xi_1$& $ka\xi_1$&$3ka\xi_1$& ${1\over 2}-a(1-3k)\xi_1$\\
$\Sigma^+$ & 
${{2ka}\over 3}(4\xi_1-\xi_2)$& ${{ka}\over 3}(4\xi_1-\xi_2)$& 
$ka(4\xi_1-\xi_2)$& ${1\over 2}-{a\over 3}(1-3k)(4\xi_1-\xi_2)$\\
$\Lambda^0$&$2ka\xi_2$& $ka\xi_2$&$3ka\xi_2$&${1\over 2}-a(1-3k)\xi_2$\\
$\Xi^0$ & ${{2ka}\over 3}(4\xi_2-\xi_1)$& ${{ka}\over 3}(4\xi_2-\xi_1)$& 
$ka(4\xi_2-\xi_1)$& ${1\over 2}-{a\over 3}(1-3k)(4\xi_2-\xi_1)$\\
\end{tabular}
\end{center}
\end{table}

\begin{table}[ht]
\begin{center}
\caption{The spin-weighted quark contents in the decuplet baryons, where
$\xi_1=1+\epsilon+f$ and $\xi_2=2\epsilon+f_s$.}
\begin{tabular}{ccccc} 
 Baryon &$\Delta u^{B^*}$ & $\Delta d^{B^*}$& $\Delta s^{B^*}$&
$\Delta\Sigma^{B^*}$ \\
\hline 
$\Delta^{++}$ & $3-3a(1+\epsilon+2f)$& $-3a$ & $-3a\epsilon$&
$3-2a(3\xi_1)$\\
$\Delta^{+}$ & $2-a(3+2\epsilon+4f)$& $1-a(3+\epsilon+2f)$ &
$-3a\epsilon$& $3-2a(3\xi_1)$\\
$\Sigma^{*+}$ & $2-a(2+3\epsilon+4f)$& $-a(2+\epsilon)$ &
$1-2a(2\epsilon+f_s)$& $3-2a(2\xi_1+\xi_2)$\\
$\Sigma^{*0}$ & $1-2a(1+\epsilon+f)$& $1-2a(1+\epsilon+f)$ &
$1-2a(2\epsilon+f_s)$& $3-2a(2\xi_1+\xi_2)$\\
$\Xi^{*0}$ & $1-a(1+3\epsilon+2f)$& $-a(1+2\epsilon)$ &
$2-a(5\epsilon+4f_s)$& $3-2a(\xi_1+2\xi_2)$\\
$\Omega^{-}$ & $-3a\epsilon$& $-3a\epsilon$ &
$3-6a(\epsilon+f_s)$& $3-2a(3\xi_2)$\\
\end{tabular}
\end{center}
\end{table}

\begin{table}[ht]
\begin{center}
\caption{The orbital angular momenta carried by quarks and antiquarks 
in the decuplet baryons, where $\xi_1=1+\epsilon+f$ and
$\xi_2=2\epsilon+f_s$.}
\begin{tabular}{ccccc} 
 Baryon &$<L_z>_q^{B^*}$ & $<L_z>_{\bar q}^{B^*}$& $<L_z>_{q+\bar
q}^{B^*}$& $<J_z>_{q+\bar q}^{B^*}$ \\
\hline 
$\Delta$&${{2ka}}(3\xi_1)$&${{ka}}(3\xi_1)$&$3ka(3\xi_1)$
&${3\over 2}-a(1-3k)(3\xi_1)$\\
$\Sigma$&${{2ka}}(2\xi_1+\xi_2)$&${{ka}}(2\xi_1+\xi_2)$&
$3ka(2\xi_1+\xi_2)$& ${3\over 2}-a(1-3k)(2\xi_1+\xi_2)$\\
$\Xi$& ${{2ka}}(\xi_1+2\xi_2)$& ${{ka}}(\xi_1+2\xi_2)$ &
$3ka(\xi_1+2\xi_2)$& ${3\over 2}-a(1-3k)(\xi_1+2\xi_2)$\\
$\Omega$&${{2ka}}(3\xi_2)$& ${{ka}}(3\xi_2)$ &
$3ka(3\xi_2)$& ${3\over 2}-a(1-3k)(3\xi_2)$\\
\end{tabular}
\end{center}
\end{table}

\end{document}